\documentclass[twoside,12pt]{article}  
\usepackage{CJK} 
\usepackage{indentfirst} 
\usepackage{bm} 
\usepackage{graphicx} 
\usepackage{amsmath} 
\usepackage{color}
\usepackage{subfigure}
\usepackage[english]{babel}
\usepackage{epstopdf}

\begin{document} 

\begin{CJK*}{GBK}{song}  



\begin{center}
\LARGE\bf Entanglement entropy for a particle coupled with its surrounding     
\end{center}

\begin{center}  
Chisanupong Puttarprom, Sikarin Yoo-Kong$^{\ddag,\dag}$*, Monsit Tanasittikosol$^{\ddag}$ and Watchara  Liewrian$^{\ddag}$
\end{center}

\begin{center}  
\begin{small} \sl
$^{\ddag}$Theoretical and Computational Physics (TCP) Group, Department of Physics,
Faculty of Science, King Mongkut's University of Technology Thonburi, Thailand, 10140.\\
*Ratchaburi Campus, King Mongkut's University of Technology Thonburi, Thailand, 10140.
\end{small}
\end{center}


\vspace*{2mm}

\begin{center}  
\begin{minipage}{15.5cm}
\begin{center}
\textbf{Abstract}
\end{center}
\parindent 20pt\small
We investigate the entanglement for a model of a particle moving in the lattice (many-body system). The interaction between the particle and the lattice is modelled using Hooke's law. The Feynman path integral approach is applied to compute the density matrix of the system. The complexity of the problem is reduced by considering two-body system (bipartite system). The spatial entanglement of ground state is studied using the linear entropy. We find that increasing the confining potential implies a large spatial separation between the two particles. Thus the interaction between the particles increases according to Hooke's law. This results in the increase in the spatial entanglement.
\end{minipage}
\end{center}

\begin{center}  
\begin{minipage}{15.5cm}{\bf {Keywords}}: bipartite system, entanglement, linear entropy, path integral
\begin{minipage}[t]{2.3cm}
\end{minipage}
\begin{minipage}[t]{13.1cm}
\end{minipage}\par\vglue8pt
{\bf PACC: }
03.65.-w,03.67.-a,03.65.Ud

\end{minipage}
\end{center}


\section{Introduction}  
Quantum theory is one of the most important branches in physics and is well-known for being counter-intuitive and its bizarre phenomena (nothing like we experience in everyday life). Quantum entanglement, which is the direct consequence of the EPR paradox \cite{1,2}, is a very unique phenomenon and has no classical counterparts. Since the discovery of quantum entanglement, a wide range of applications has been proposed, i.e., quantum cryptography, quantum teleportation and superdense coding \cite{3,4,5,6}. These applications indicate that entanglement could be exploited to perform the impossible tasks which cannot be done by classical computers. Another new branch of physics, which is called quantum information and computing, has been born by joining quantum physics and theoretical computer science. In this respect many physical systems have been proposed as promising hardwares to perform quantum information processors \cite{7,8,9,10,11,12,13,14}.  For the solid state systems, the main difficulty of studying is the complexity accounting to many-body configuration. The approximation methods must be introduced to study interested physical properties, see \cite{15,16}.

One of the most interesting systems in the context of solid state physics is the polaron which was initiated by Landau \cite{17}, Pekar \cite{18} and Frohlich \cite{19}. Later, Feynman \cite{20} offered a new way, called path integrals, to study the physical properties, e.g., the ground state energy and the effective mass \cite{21} and references therein]. The main object in this method is the propagator which is equivalent to the density matrix under the transformation from real time to imaginary time \cite{22,23}. Unfortunately, the propagator for the polaron could not be exactly computed and the approximation was needed. The trial system was introduced to replace the actual system and the Lagrangian is given by \cite{24,32,33,34}
\begin{eqnarray}\label{L1}
\mathcal{L}_{\boldsymbol{r},\boldsymbol{R}}&=&\frac{m}{2}\dot {\boldsymbol{r}}^2(\tau)- V(\boldsymbol{r}) + \sum_{i=1}^N\left( \frac{M}{2}\dot {\boldsymbol{R}}_i^2(\tau)-\frac{M\Omega^2}{2}\boldsymbol R_i^2(\tau) \right)\nonumber\\
&&-\frac{\kappa}{2}\sum_{i=1}^N|\boldsymbol r(\tau)-\boldsymbol R_i(\tau)|^2\;,
\end{eqnarray}
where $\boldsymbol{r}$ and $\boldsymbol{R}_i$ are the position vectors of the particle with mass $m$ and the particles with mass $M$, respectively. The $\Omega$ is the vibration frequency (This frequency can be connected with the ambient temperature of the system) of the particles with mass $M$, $\kappa$ is the coupling constant and 
\[
V(\boldsymbol{r})=\frac{m}{2}\left[\omega_x^2 x^2 + \omega_y^2 y^2 + \omega_z^2 z^2  \right] \;,
\]
is the confining potential: $\omega_z = 0$ for quantum wires and $\omega_y = \omega_z = \omega$ for quantum dots. Interestingly, the system in Eq. \eqref{L1} has long been studied, \emph{but no one has ever investigated the quantum correlation, quantum entanglement, in such system, namely between the particle and the lattice}. To this contribution, we study the entanglement of the ground-state between the particle and the lattice in Eq. \eqref{L1} with some simplification (which will be described in the next section). We will focus on the entanglement generated by the spatial degrees of freedom. The linear entropy will be used to measure the amount of entanglement.
\begin{figure}[h!]
  \centering 
   \includegraphics[width=0.50\textwidth]{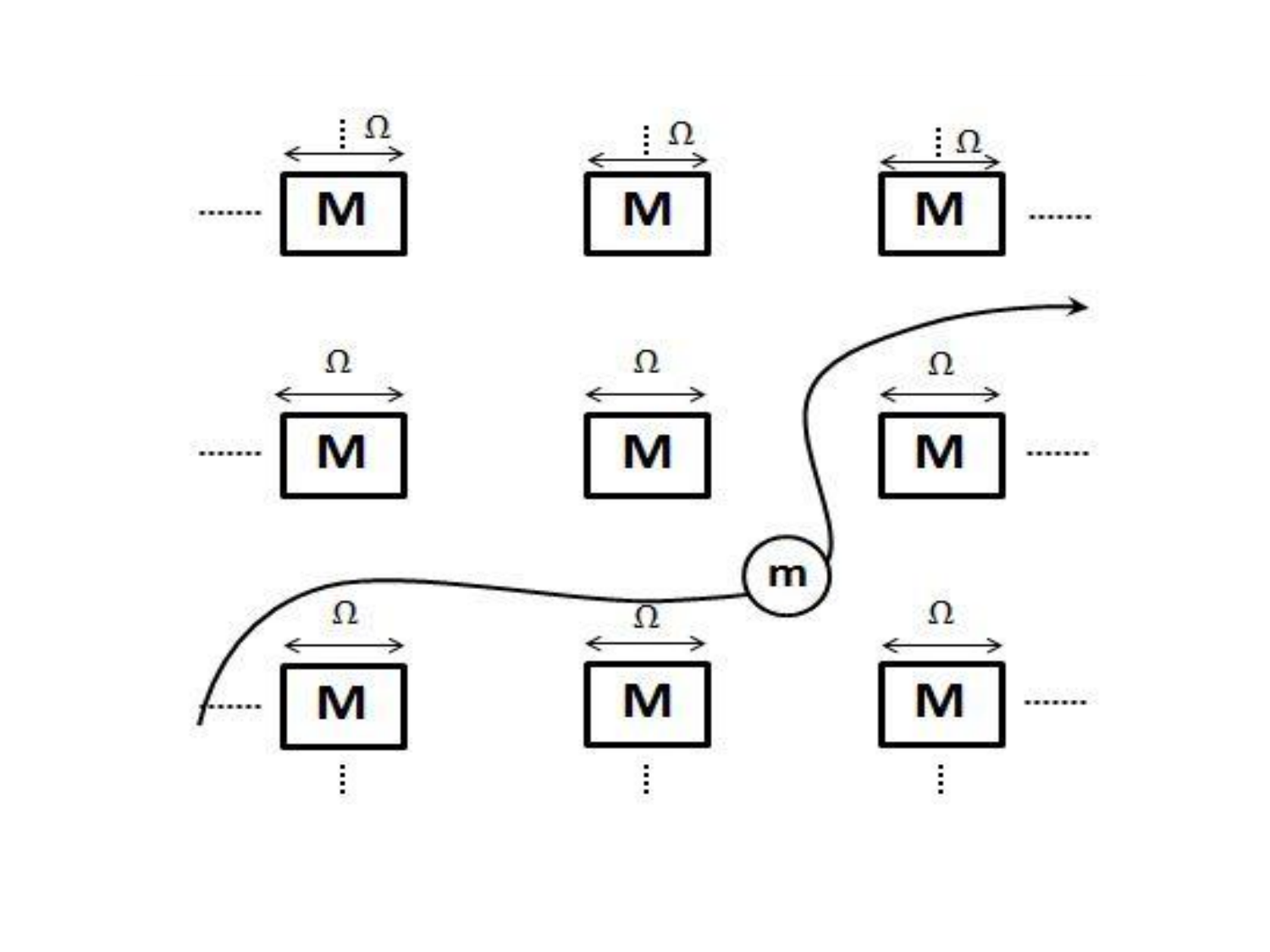} 
  \caption{{\small{The Lagrangian in Eq \eqref{L1} describes a particle of mass $m$ moving in the lattice which is constituted from the particles of mass $M$. The interaction between the particle and the lattice is modelled by Hook's law. The external potential $V(\boldsymbol{r})$ can be represented the shape of the matter as we mentioned in the main text.}}} 
   \label{fig:2}
\end{figure}
\\

The organisation of this paper is the following. In section \ref{sec:2} ~~ we introduce the toy model: a coupled harmonic oscillator and compute the density matrix. Then, the linear entropy will be briefly discussed and will be used to quantify the spatial entanglement of the system in section {\ref{sec:3}} ~~ The last section will be devoted to summary and discussion.

\section{The system}\label{sec:2}
According to the complexity of the system in Eq. \eqref{L1}, quantum many-body approach is needed to study the problem. Practically, the wave function of the particle interacting with the lattice constitutes a Hilbert space $\mathcal{H}: |\Psi > \in \mathcal{H} = \mathcal{H}_{particle} \otimes  \mathcal{H}_{lattice}$. The density matrix is given by
\begin{eqnarray}\label{H1}
\rho = |\Psi ><\Psi | \;.
\end{eqnarray}
The reduced density matrix is obtained by tracing out the environment
\begin{eqnarray}\label{H1}
\rho = |\Psi ><\Psi | \;,
\end{eqnarray}
which is described part of a system which is the particle in this case. Alternatively, the Feynman's path integral approach provides another accessible way to tackle the problem and the exact form of the propagator \cite{26}, which can indeed be used to find the wave function. Furthermore, the density matrix of the system can be obtained directly from the propagator under the transformation from the real time variable to the complex time variable.

To compute the density matrix of the system in Eq. \eqref{L1}, we start to consider the propagator of the composite system which is given by
\begin{eqnarray}{\label{propagator}}
\mathcal{K}_{\boldsymbol{r},\boldsymbol{R}}&=&\prod_{i=1}^N\mathcal{K}(\boldsymbol{r}(t),\boldsymbol{R}_i(t),t;\boldsymbol{r}(0),\boldsymbol{R}_i(0),0) \nonumber \\
&=&{\int_{\boldsymbol{r}(0)}^{\boldsymbol{r}(t)}} \mathcal{D}(\boldsymbol{r}) \prod_{i=1}^N{\int_{\boldsymbol{R}_i(0)}^{\boldsymbol{R}_i(t)}}\mathcal{D}(\boldsymbol{R}_i){\rm{e}}^{{\frac{\rm_{i}}{\hbar }}S}\;,
\end{eqnarray}
where the action is given by
\begin{eqnarray}
S&=&{\int_{0}^{t}}{\rm{d}\tau }\mathcal{L}_{\boldsymbol{r},\boldsymbol{R}}\;.
\end{eqnarray}
We perform the transformation from the real time to the imaginary time: $t\rightarrow -{\rm i}\beta$ where $\beta = 1/k_{\rm B}T$ with $\hbar =1$. The parameter $T$ is the temperature and $k_{\rm B}$ is the Boltzmann constant~\cite{27}. Under this transformation,The propagator Eq.~{\eqref{propagator}} becomes the density matrix.
\begin{eqnarray}
\rho_{\boldsymbol{r},\boldsymbol{R}}(\boldsymbol{r}(\beta),\boldsymbol{R}(\beta),\beta;\boldsymbol{r}(0),\boldsymbol{R}(0),0)&=&{\int_{\boldsymbol{r}(0)}^{\boldsymbol{r}(\beta)}} \mathcal{D}(\boldsymbol{r}) \prod_{i=1}^N{\int_{\boldsymbol{R}_i(0)}^{\boldsymbol{R}_i(\beta)}}\mathcal{D}(\boldsymbol{R}_i){\rm{e}}^{-S_{\beta }}\;,
\end{eqnarray}
where
\begin{eqnarray}
{S}_{\beta }&=&{\int_{0}^{\beta}}{\rm{d}}\tau \mathcal{L}_{\boldsymbol{r},\boldsymbol{R}}\;.
\end{eqnarray}
The reduced density matrix can be obtained by tracing out the variable $\boldsymbol{R}_i$ of particles with mass $M$ that constitute the lattice\cite{27,28,29}.
\begin{eqnarray}\label{S1}
\rho_{\boldsymbol{r}}(\boldsymbol{r}(\beta ),\boldsymbol{r}(0))&=& {\rm{Tr}}_{\boldsymbol{R}}\rho_{\boldsymbol{r},\boldsymbol{R}} 
={\int_{\boldsymbol{r}(0)}^{\boldsymbol{r}(\beta )}} \mathcal{D}(\boldsymbol{r})\prod_{i=1}^N \int d\boldsymbol{R}_i{\int_{\boldsymbol{R}_i}^{\boldsymbol{R}_i}}\mathcal{D}(\boldsymbol{R}_i)e^{-S_{\beta }}  \nonumber \\
&=&\left[\frac{1}{2\sinh(\Omega_{\rm{eff}}\beta/2)}\right]^{dN}
{\int_{\boldsymbol{r}(0)}^{\boldsymbol{r}(\beta )}} \mathcal{D}(\boldsymbol{r})e^{-S_{\beta }^\prime}\;,
\end{eqnarray}
where
\begin{eqnarray}
S^\prime_{\beta}&=&\int_{0}^{\beta}d\tau\left(\frac{m}{2}\dot {\boldsymbol{r}}^2(\tau)- V(\boldsymbol{r}(\tau)) 
-\frac{N\kappa}{2}\boldsymbol{r}^2(\tau) \right)\nonumber\\
&&-\frac{N\kappa^2}{4M \Omega_{\rm{eff}}}\int_{0}^{\beta}d\tau\int_{0}^{\beta}d\sigma\frac{\cosh(\Omega_{\rm{eff}}|\tau-\sigma|-\beta/2)}{\sinh(\Omega_{\rm{eff}}\beta/2)}\boldsymbol{r}(\tau)\boldsymbol{r}(\sigma)\;,\\
\Omega_{\rm{eff}}^2&=&\Omega^2+\frac{N\kappa}{M}\;.
\end{eqnarray}
\begin{figure}[h!]
  \centering 
   \includegraphics[width=0.50\textwidth]{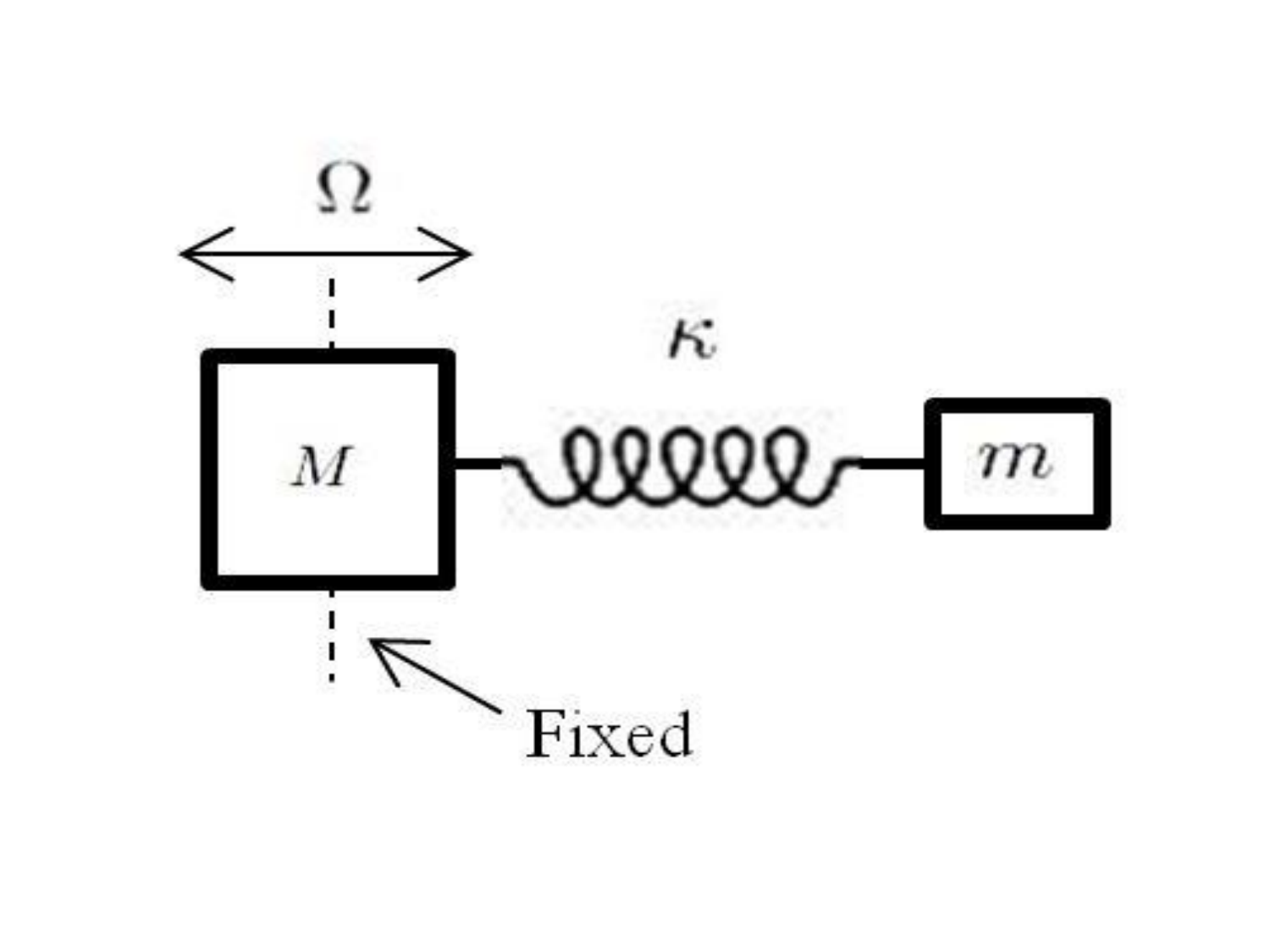} 
  \caption{\small{The system of a coupled harmonic oscillator. The particle of mass $M$ is fixed to its equilibrium while the particle of mass $m$ is free to move in one dimension (horizontal line). The connection between these two particles is modelled through the spring interaction.}} 
   \label{fig:2}
\end{figure}
The variable $d$ in Eq. \eqref{S1} is the number of dimensions. To reduce the complexity of the problem in Eq. \eqref{S1}, we will focus on a bipartite system consisting of one particle with mass $m$ and one particle with mass $M$ $(N=1)$ in one dimension {\color{blue}$(d=1)$. This simple system can be treated as a coupled harmonic oscillator in one dimension as shown in Fig.~\ref{fig:2}. The particle with mass $M$ vibrates around its equilibrium with frequency $\Omega$. The particle with mass $m$ is trapped in the harmonic potential $V(x)=m\omega^2x^2/2$ with frequency $\omega $ and also interacts with mass $M$. The interaction between these two particles is modelled by Hooke's law and $\kappa$ denotes the strength of the interaction. Now the density matrix in Eq. \eqref{S1} becomes
\begin{eqnarray}\label{S3}
\rho_{x}(x(\beta ),x(0))
&=&\left[\frac{1}{2\sinh(\Omega_{\rm{eff}}\beta/2)}\right]{\int_{x(0)}^{x(\beta )}} \mathcal{D}(x)e^{-S_{\beta }^\prime}\;,
\end{eqnarray}
where
\begin{eqnarray}
S^\prime_{\beta}&=&\int_{0}^{\beta}d\tau\left(\frac{m}{2}\dot {x}^2(\tau)- \frac{m\omega^2}{2}x^2(\tau)
-\frac{\kappa}{2}x^2(\tau) \right)\nonumber\\
&&-\frac{\kappa^2}{4M\Omega_{\rm{eff}}}\int_{0}^{\beta}d\tau\int_{0}^{\beta}d\sigma\frac{\cosh(\Omega_{\rm{eff}}|\tau-\sigma|-\beta/2)}{\sinh(\Omega_{\rm{eff}}\beta/2)}x(\tau)x(\sigma)\;.
\end{eqnarray}
The path integral in Eq. \eqref{S3} can be found \cite{23,31}. The path integral in Eq. \eqref{S3} can be solved exactly \cite{30}
\begin{eqnarray}\label{S4}
\rho_{x}(x(\beta ),x(0))&=& \left({\frac{m}{8 \pi \hbar \triangle (0)\sinh(\Omega_{\rm{eff}}\beta/2)}}\right)^{\frac{1}{2}} \left(\frac{\sinh({\frac{\Omega_{\rm{eff}} \beta}{2}})}{{\sinh({\frac{z_{+} \beta }{2}})}{\sinh({\frac{z_{-} \beta }{2}})}}\right)\nonumber \\
&&\times {\rm{e}}^{[{{\frac{m \ddot{\bigtriangleup}(0) }{4}}(x(\beta )-x(0))^2}-{{\frac{m}{4 \bigtriangleup (0)}}(x(\beta )+x(0))^2}]}\;,
\end{eqnarray}
where
\begin{subequations}
\begin{eqnarray}
\bigtriangleup (\tau )&=&{\frac{z_{+}^2-{{\Omega}^2}_{\rm{eff}}}{z_{+}^2-z_{-}^2}} {\frac{\cosh({z_{+}}(\tau -{\frac{\beta }{2}}))}{{z_{+} \sinh({\frac{z_{+} \beta }{2}})}}} 
+{\frac{z_{-}^2-{{\Omega}^2}_{\rm{eff}}}{z_{-}^2-z_{+}^2}} {\frac{\cosh({z_{-}}(\tau -{\frac{\beta }{2}}))}{{z_{-} \sinh({\frac{z_{-} \beta }{2}})}}} \;,\\
2z^2_{\pm }&=&\frac{1}{2}\left( \frac{\kappa}{m}+\omega^2+\Omega^2_{\rm{eff}}\right) \nonumber\\
&&\pm\frac{1}{2} \sqrt{\frac{4m+M}{m^2M}\kappa^2+(\omega^2-\Omega^2_{\rm{eff}})\left(\frac{2\kappa}{m}+\omega^2-\Omega^2_{\rm{eff}}\right)}\;.
\end{eqnarray}
\end{subequations}
The reduced density matrix derived in Eq. \eqref{S4} will play an important role in the next section as a main equation to study the entanglement of the system.
\section{Entanglement and linear entropy}\label{sec:3}

The entanglement generated by the continuous spatial degree of freedom is expected to depend on the interplay between the strength of the confining potential and interaction. Then we are interested in investigating how the spatial entanglement changes with the system parameters by considering at the linear entropy of the reduced density matrix:
\begin{eqnarray}\label{S5}
\mathit{S}_{\rm_{L}}&=&1-{\rm{Tr}}\rho_{\rm{red}} ^2\;,\;\;\;\;0\leq \mathit{S}_{\rm_{L}}\leq 1\;,
\end{eqnarray}
which is used to measure the entanglement for a pure state. The linear entropy Eq. \eqref{S5} is bounded by the factor $k/(k-1)$, where $k$ is the dimension of the state. In this work, we are interested in the spatial entanglement between the two particles, hence $k$ is infinite. Then the linear entropy is bound to unity. The system would be completely disentangled when $\mathit{S}_{\rm_{L}} = 0$. On the other hand, the system would be in the maximally entangled state if  $\mathit{S}_{\rm_{L}} = 1$. For the continuous variable case, ${\rho}_{\rm_{red}}^2$ can be computed by
\begin{eqnarray}
\rho_{\rm{red}} ^2 (x,x')=\int {\rm{d}}x'' \rho_{\rm{red}}(x,x'') \rho_{\rm{red}}(x'',x')\;, 
\end{eqnarray}
and
\begin{eqnarray}
{\rm{Tr}}\rho_{\rm{red}}^2=\int {\rm{d}}x \rho_{\rm{red}}^2(x,x)\;.
\end{eqnarray}
\begin{figure}[h!]
\centering

   \includegraphics[width =.8\textwidth] {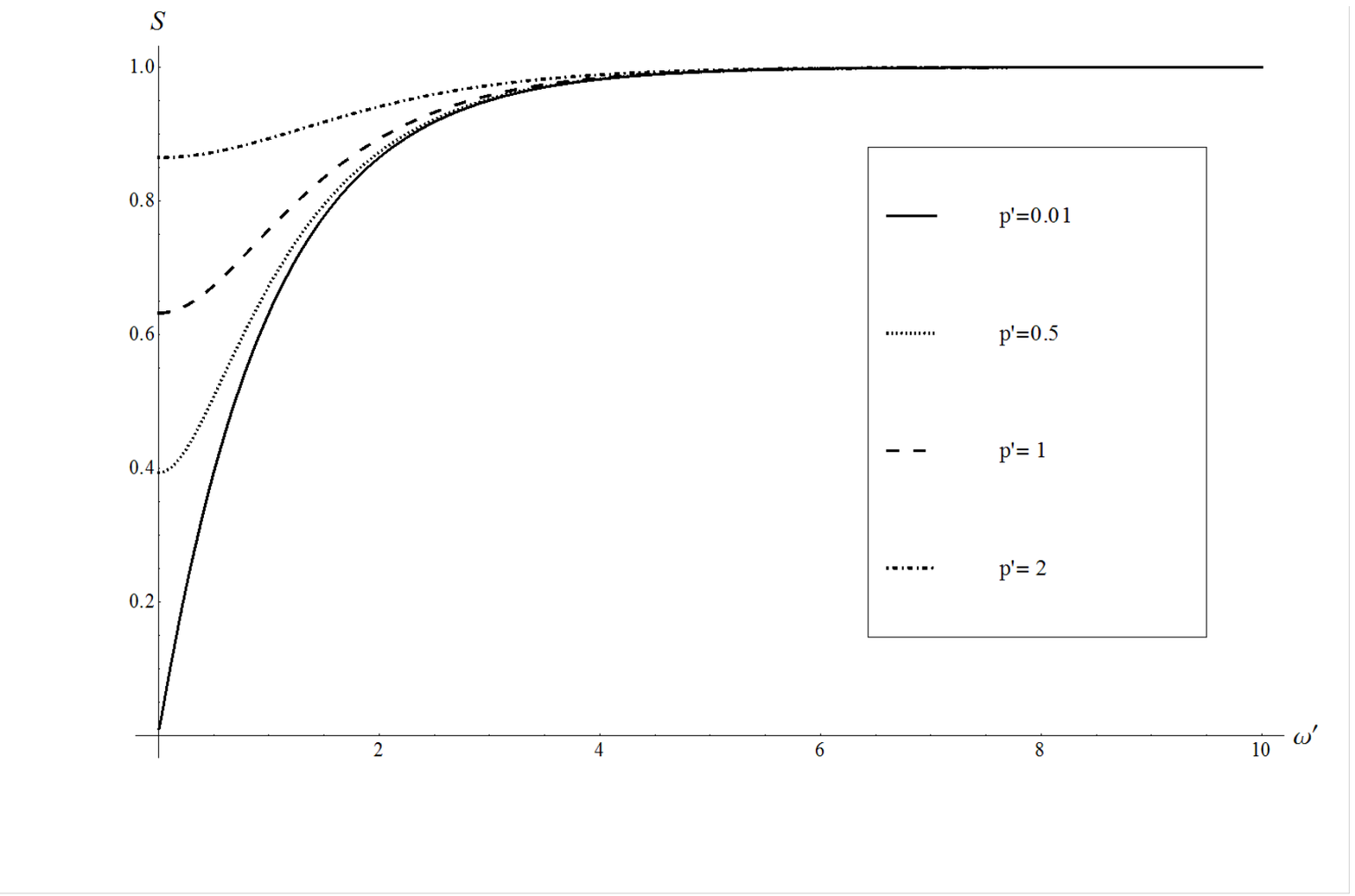}\label{ee}

\caption{The numerical results of the linear entropy. This figure shows the variation of the linear entropy with respect to the confining potential $\omega' = \omega / p$ where $p=\kappa / m$.}
\label{fig:4}
\end{figure}
It is true that the linear entropy is the first order approximate of the von Neumann entropy. However, it has been used widely in many systems and also has been sufficiently proved as one of the tools to measure the entanglement \cite{16,25}. The advantage of the linear entropy is that there is no need to diagonalise the density matrix. Then the linear entropy is much simpler comparing with the von Neumann entropy. Taking $\beta\rightarrow \infty$ in Eq. \eqref{S4}, the system will go to the ground state. We now find that the linear entropy of our system is
\begin{eqnarray}{\label{trp}}
\mathit{S}_{\rm_{L}}=1-{\rm{Tr}}\left[\lim_{\beta\rightarrow \infty}\rho_{x}\right]^2\;.
\end{eqnarray}
Figure~3 shows the numerical results of how $\mathit{S}_{\rm_{L}}$ changes with respect to $\omega^\prime$ in the limit $M>>m$. It is clear that the spatial entanglement increases when the confining potential $\omega$ increases. This is because the spatial separation between $m$ and $M$ increases resulting from the high confining potential and hence $\mathit{S}_{\rm_{L}}$ increases \cite{15,16}. 


\section{Summarising discussion}
We have studied the spatial entanglement through the linear entropy for the coupled harmonic oscillator. 
The reduced density matrix of the system has been exactly computed by using Feynman path integral approach and has been changed from the time domain to temperature domain through the imaginary time transformation. At the ground state corresponding to the limit $\beta\rightarrow \infty$, the entanglement entropy increases with increasing of the confining potential. This result implies that entanglement is proportion to the distance between two particles. This seems to suggest that the entanglement between these two particles can be tuned by varying the strength of the trap. For the case of arbitrary $N$, the complexity of system increases, but the density matrix in this paper can be used as a basic ingredient.
 
 We believe that in this preliminary paper we provide an alternative way to study the entanglement entropy through the computation of the density matrix according to Feynman path integral, apart from the standard method \cite{4, 6, 16}}. This method could be possibly applied to study the entanglement in  solid state physics such as polaron, exiton and plasmon. Furthermore, the question we intend to investigate next is that whether entanglement between the particle and its surrounding affects other physical properties of the quasiparticle, e.g., effective mass, conductivity and stability.
\section*{Acknowledgements}
S. Yoo-Kong gratefully acknowledges the support from the Research Projects of Faculty of Science, King Mongkut's University of Technology Thonburi. 

\vspace*{2mm}


\end{CJK*}  
\end{document}